\documentclass[reprint]{revtex4-2}
\usepackage{dcolumn}
\usepackage[T1]{fontenc}
\usepackage{amsmath, amssymb}
\usepackage{graphicx}
\usepackage{caption}
\usepackage{geometry}
\usepackage{setspace}
\usepackage{hyperref}
\usepackage{float}
\usepackage{enumitem}
\begin{document}

\preprint{APS/123-QED}

\title{\textbf{Ant Colony Optimization for Density Functionals in Strongly Correlated Systems}}

\author{%
Gabriel M. Tonin\textsuperscript{}, 
Tatiana Pauletti\textsuperscript{}, 
Ramiro M. Dos Santos\textsuperscript{},
Vivian V. França\textsuperscript{*}\\
\small\textsuperscript{}São Paulo State University (UNESP), Institute of Chemistry, 14800-090, Araraquara, São Paulo, Brazil\\
\small *Corresponding authors: vivian.franca@unesp.br\\
\singlespacing
}

\date{\today}

\begin{abstract}
\noindent The Ant Colony Optimization (ACO) algorithm is a nature-inspired metaheuristic method used for optimization problems. Although not a machine learning method per se, ACO is often employed alongside machine learning models to enhance performance through optimization. We adapt an ACO algorithm to optimize the so-called FVC density functional for the ground-state energy of strongly correlated systems. We find the parameter configurations that maximize optimization efficiency, while reducing the mean relative error ($MRE$) of the ACO functional. We then analyze the algorithm's performance across different dimensionalities ($1D-5D$), which are related to the number of parameters to be optimized within the FVC functional. Our results indicate that $15$ ants with a pheromone evaporation rate superior to $0.2$ are sufficient to minimize the $MRE$ for a vast regime of parameters of the strongly-correlated system --- interaction, particle density and spin magnetization. While the optimizations $1D$, $2D$, and $4D$ yield $1.5\%< MRE< 2.7\%$, the $3D$ and $5D$ optimizations lower the $MRE$ to $\sim0.8\%$, reflecting a $67\%$ error reduction compared to the original FVC functional ($MRE = 2.4\%$). As simulation time grows almost linearly with dimension, our results highlight the potential of ant colony algorithms for density-functional problems, combining effectiveness with low computational cost.


\end{abstract}

\maketitle


\section{Introduction}
\label{sec:introduction}
Recently, several artificial intelligence algorithms have been adapted to problems in different scientific areas, in particular in physics and chemistry \cite{Prezhdo_9656_2020,Karthikeyan_0973_2021,Chen_1463_2025}. For example, machine learning has contributed to the development of new molecules \cite{Tropsha_1421_2023,Barcin_136668_2024,Qi_903_2024} and also to the understanding of several properties of many-body systems \cite{Omranpour_1616_2025,Pikalova_9_2024,Andolina_1421_2024}. Approaches based on biological systems have gained space in this context, such as neural networks based on brain cell function, which have contributed to solving complex problems in computational physics and mathematics\cite{Custdio_s41598_2019,Dornheim_1097_2023,Habring_47_2024}. 

Another example inspired by biological systems is the metaheuristic Ant Colony Optimization (ACO) \cite{Dorigo_243_2005, Dorigo_250_2003}, based on how ants learn the minimal path between the nest and food. While ACO is not a machine learning method in the strict sense, it is frequently integrated into machine learning workflows to optimize performance\cite{Chen_3308_2007,Chen_1222_2025}. Ants follow paths with high concentration of pheromone \cite{Dorigo_28_2004} $-$ a chemical substance they produce and deposit along the way $-$ thus guiding the colony to the food source and back to the nest along the optimal path. This can be adapted for a computational algorithm, introducing the concept of the pheromone update to find solution to mathematical problems. Recent studies applying the ACO for two variables ($2D$ optimization) demonstrate the applicability of ACO in a fast medicine dispensing system \cite{Jin_168781401771370_2017} and mobile robot trajectory planning \cite{Li_6756_2020}. Despite its potential, few studies have investigated ACO's effectiveness in high-dimensional optimization problems \cite{Riabko_012001_2022, Shmygelska_1471_2005,Wu_7464_2021}.

In the context of many-electron systems, one of the main goals is to determine the ground-state energy \cite{Goshen_2337_2024}. Several approaches, such as Hartree-Fock methods \cite{Slater_358_1951,Valatin_1012_1961}, configuration interactions (CI) \cite{Witzorky_2469_2024,Nys_2041_2024}, and density functional theory (DFT) \cite{vanMourik_20120488_2014,Kohn_1253_1999}, apply the variational principle \cite{Wu_296_2024,Wang_1079_2022} to solve self-consistently this problem.   We focus particularly on approaches to obtaining the ground-state properties of systems described by the one-dimensional Hubbard model  \cite{Irkhin_2135_2022,Hubbard_238_1963}. Despite its simplicity, the model captures many important phenomena, such as the Mott metal-insulator transition \cite{Mott_677_1968,Canella_2469_2021,Pauletti_0948_2024}, Anderson localization \cite{Anderson_1492_1958}, and conventional and exotic superfluidity \cite{Arisa_214522_2020,Fulde_A550_1964,Larkin_762_1965}. Analytical approximations to the ground-state energy of the homogeneous Hubbard model have been proposed, initially for spin-independent problems \cite{Lima_146402_2023}, while the so-called FVC functional \cite{Franca_073021_2012} extends the approach to magnetized systems. They provide density functionals parametrizations for the numerical Bethe-Ansatz  solution \cite{Lieb_1_2023}, enabling the description of inhomogeneous Hubbard chains via a local-density approximation (LDA) \cite{Lima_1286_2002,Xianlong_1550_2006,Palamara_062203_2024,Zhang_2025_013184}. 

In this work, we introduce the ant colony algorithm as a novel approach to optimize density functionals. We assess the ACO potential by applying it to optimize the FVC functional~\cite{Franca_073021_2012}. We investigate the optimization across different dimensions ($1D - 5D$), evaluating the performance via the global mean relative error ($MRE$) for a vast regime of interactions, particle densities and magnetizations. The $3D$ case, in particular, exhibited the highest optimization efficiency, achieving a $MRE\sim0.8\%$, which corresponds to a $67\%$ reduction of FVC error ($MRE=2.4\%$), thus offering an effective trade-off between accuracy and computational cost. Our results not only validate the applicability of ACO to high-dimensional problems, but also reveal that ant colony algorithms are highly effective for optimizing density functionals.

\section{Ant Colony Optimization}

Ant Colony Optimization (ACO) is a probabilistic metaheuristic method inspired by the way ants search for food and navigate their environment \cite{Mostafaie_104850_2020}. Ants deposit pheromones along their paths, creating trails that influence the movement of other ants. By self-organizing, colonies are able to collaboratively find optimal paths, a mechanism that inspires the design of ACO algorithms to solve discrete optimization tasks. The method is based on two core concepts: exploration of new paths and pheromone-guided decision-making. 

New paths are explored through the probability that an ant $k$ moves from a state $x$ to a state $y$, given by \cite{Vittorio_1_2024}:
\begin{equation}
p^k_{xy} = \frac{\tau_{xy}^\lambda \eta_{xy}^\omega}{\displaystyle \sum_{z \in J^k} \tau_{xz}^\lambda \eta_{xz}^\omega}~,
\end{equation}
\noindent where $\tau_{xy}$ represents the level of pheromone on the path $x\rightarrow y$, and $\eta_{xy}$ denotes the heuristic attractiveness, often defined as the inverse of the distance between nodes $x$ and $y$. The parameters $\lambda$ and $\omega$ control the relative influence of pheromone levels and heuristic information, respectively\cite{Dorigo_243_2005}.  Heuristic information refers to problem-specific knowledge used to guide the search process toward promising solutions. In ACO, heuristic information usually manifests as a function that evaluates the desirability of moving between nodes, using criteria such as distance, cost, or other relevant metrics. A higher $\lambda$ value makes ants more likely to follow existing pheromone trails, while a higher $\omega$ value increases the influence of heuristic knowledge, such as shorter distances or lower costs. The denominator normalizes the probabilities in all unvisited nodes in the set $J^k$\cite{Dorigo_243_2005}.

\begin{figure}[htp!]
\centering
\includegraphics[width=1\columnwidth]{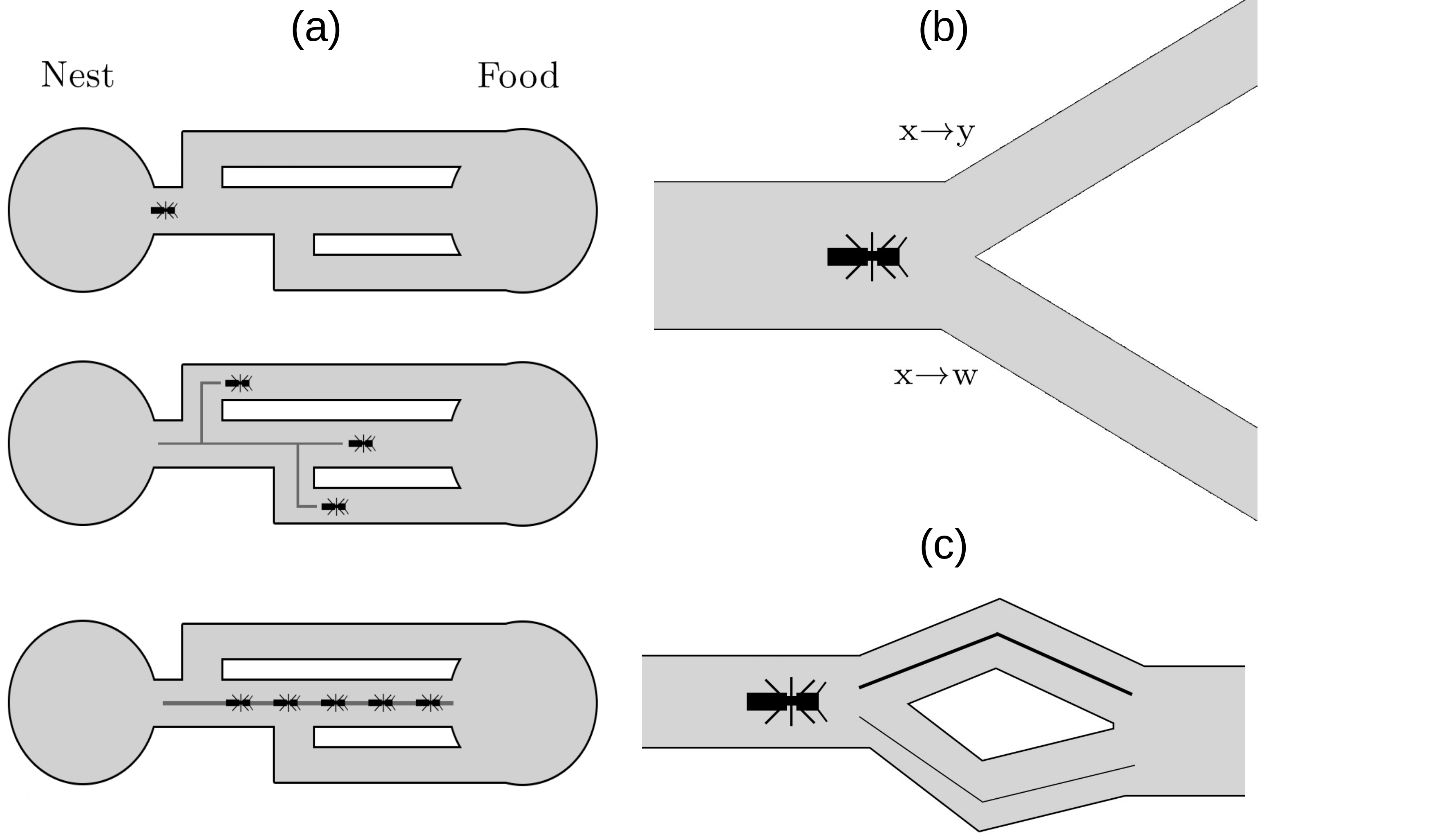}

\vspace{0.5cm} 

\includegraphics[width=1\columnwidth]{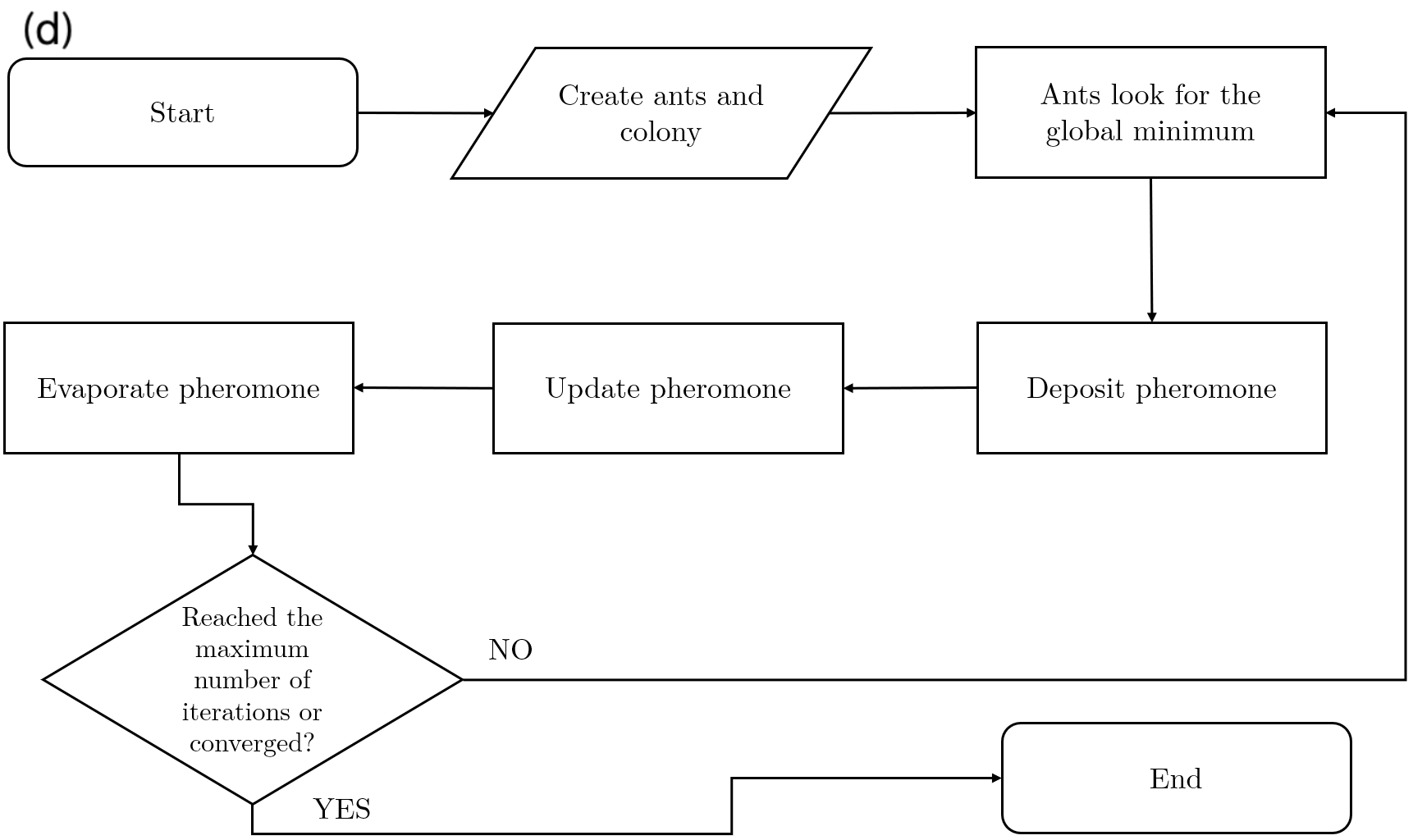}

\caption{Illustration of ant behavior during optimization: (a) Ants searching for paths, (b) Multiple possible routes, and (c) Pheromone-based optimization with stronger trails (thicker lines) indicating better paths. (d) Flowchart of the ACO-based optimization algorithm for functional minimization.}
\label{fig:antsbehavior}
\end{figure}

After each iteration, pheromone levels are updated as \cite{Vittorio_1_2024}:

\begin{equation}
\tau_{xy} \leftarrow \sum_{k=1}^m \Delta \tau^k_{xy}
+ (1-\rho)\tau_{xy} ~.
\label{eq:feromonio}
\end{equation}
\noindent The term $\sum_{k=1}^m \Delta \tau^k_{xy}$ represents the pheromone deposits from the $m$ ants, with each deposit proportional to the quality of the corresponding solution, defined as:

\begin{equation}
\label{eq:quality}
    \Delta \tau^k_{xy} =
    \begin{cases} 
    \frac{Q}{L_{k}}, & \text{if ant $k$ chooses path } x \rightarrow y, \\ 
    0, & \text{otherwise}.
    \end{cases}
\end{equation}
where $Q$ scales the amount of pheromone deposited within a certain path $L_k$ \cite{Riabko_012001_2022}. The quality of the solution is then proportional to the pheromone concentration per unit of length, which increases by either increasing $Q$ or decreasing $L_k$. This approach ensures that more efficient solutions receive higher pheromone increments, reinforcing their importance over time and guiding subsequent ants. The term $(1-\rho)\tau_{xy}$ in Eq.(\ref{eq:feromonio}) represents the evaporation of the pheromone, where $\rho$ is the evaporation rate. This mechanism prevents older, suboptimal paths from dominating the search space by gradually reducing their influence. 

Figure \ref{fig:antsbehavior} illustrates the effect of pheromones and the optimization process. Ants traverse the space by evaluating parameter combinations, guided by pheromone trails and heuristic factors. The algorithm iteratively updates the pheromone levels and refines the solutions, progressively improving them to ensure convergence towards the global minimum. In all calculations we have considered a total of 1000 iterations.

\section{Application to Functional Minimization}

Our target functional, derived from the FVC parametrization 
\cite{Franca_073021_2012}, is given by:
\begin{eqnarray}
e^{FVC}(n, m, U) &=& -\frac{2\beta(n, m, U)}{\pi} \sin \left( \frac{\pi n}{\beta(n, m, U)} \right) \notag\cos \left( \frac{\pi m}{\gamma(n, m, U)} \right)~.
\end{eqnarray}
Here $n=n_\uparrow+n_\downarrow$ is the electronic density, $m=n_\uparrow-n_\downarrow$ represents the magnetization and $U$ is the Coulomb interaction (in units of the hopping term $t$, set as $t=1$). The $\beta(n,m,U)$ and $\gamma(n,m,U)$ functions are defined as
\begin{equation}
\beta(n, m, U)= \beta(U)^{\alpha(n, m, U)},
\label{eq:beta}
\end{equation}
\begin{equation}
\alpha(n, m, U)= \left[ \frac{n^{P1} - m^{P1}}{n^{P2}} \right]^{U^{P3}},
\label{eq:alpha} 
\end{equation}
\begin{equation}
\gamma(n, m, U)= 2 \exp \left[ \frac{U^{P4}}{1 - \left(\frac{m}{n}\right)^{P5}} \right],
\label{eq:gama}
\end{equation}
with $\beta(U)$ obtained by solving  
\begin{equation}
\frac{-2\beta(U)}{\pi}\sin\left(\frac{\pi}{\beta(U)}\right) = -4 \int_0^\infty dx\frac{J_0(x)J_1(x)}{x(1+e^{Ux/2})},
\label{eq:b}
\end{equation}
where $J_0(x),J_1(x)$ are Bessel functions.

Dimensionality here then refers to the number of parameters $P's$ effectively used during the ACO optimization, while the others are kept fixed on their original FVC values ($P_1=2$, $P_2=15/8$, $P_3=1/3$, $P_4=1/2$, $P_5=3/2$). Thus for $1D$ optimization we have five possibilities (one for each $P$), while for $5D$ we have only one possible optimization, containing all $P's$. For $2D$, $3D$, and $4D$, we report only the combinations of $P$ parameters that yield reasonable outcomes during the optimization process.   

The aim of the optimization is to provide a more accurate analytical expression for the FVC functional. Thus we define a cost function $fval$ to be minimized within the ACO. In this case, we set $L_k=Bestfval$ in the pheromone update equation, Eq.(\ref{eq:quality}), where $Bestfval$ represents the best objective function found so far. As $Bestfval$ decreases, the quality of the solution improves, demonstrating the efficiency of the optimization. Another key indicator of optimization performance is the mean relative error ($MRE$) between the optimized energy functional $E$ and the exact numerical solution $e$, obtained via Bethe-Ansatz numerical solution \cite{Lieb_1_2023},
$MRE(\%) = 100/M\sum|E(n,m,U) - e(n,m,U)|/e(n,m,U)$.
The average is performed over $M=280,000$ data, exploring a vast range of the Hubbard parameters: $U=\{0.8,2.0,4.0,6.0,8.0,10.0,12.0\}$, for each $U$ we consider 200 values of density, $0\leq n\leq 1$, and for each $n$ we calculate 200 values of magnetization, $0\leq m \leq n$.

\section{Results and Discussion}\label{sec3}

\begin{figure}[htp!]
    \centering
     \includegraphics[width=1\linewidth]{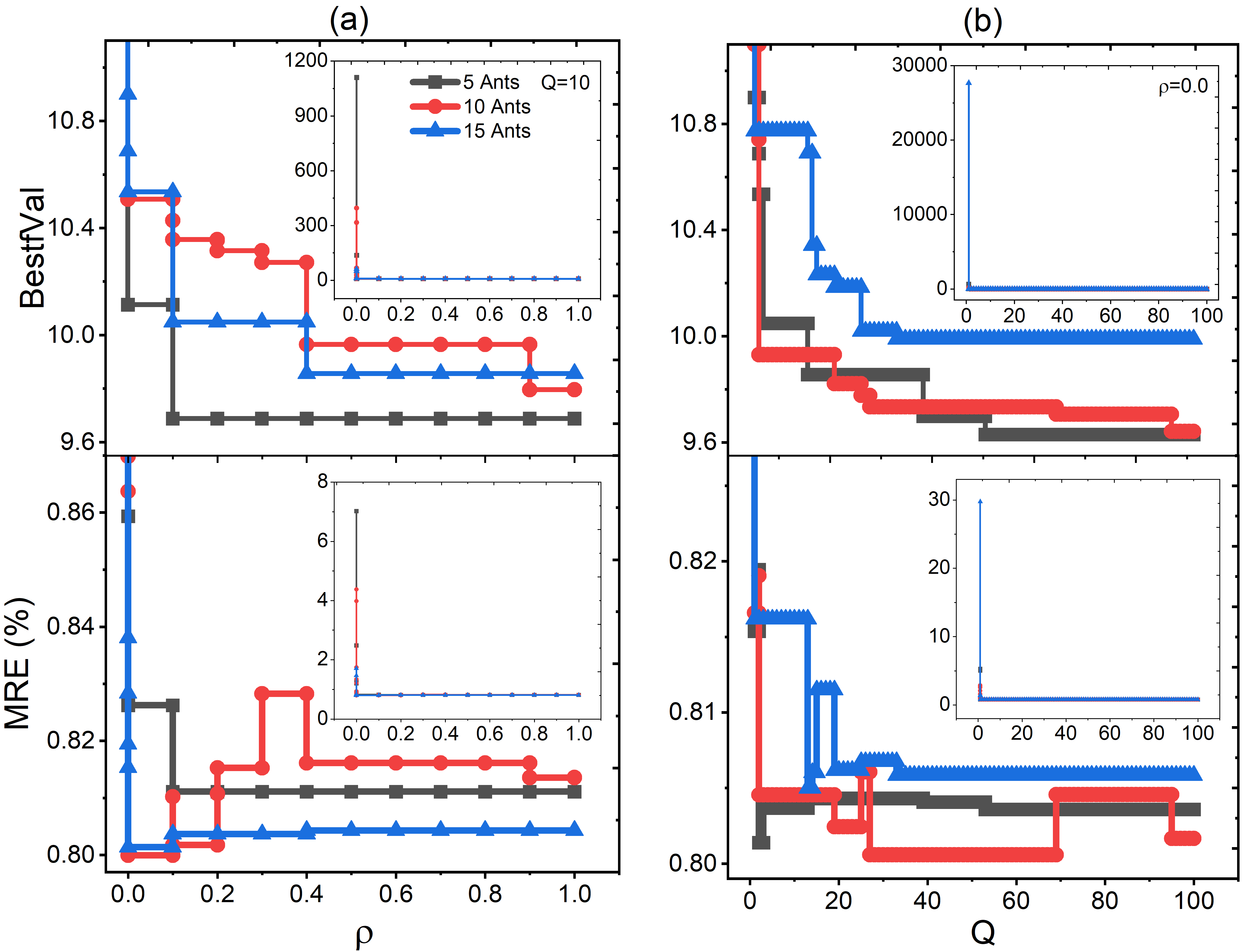}
    \caption{Analysis of the optimization performance, via $Bestfval$ and mean relative error $MRE$, as a function of (a) the pheromone evaporation rate $\rho$  and (b) the quality factor $Q$, for three numbers of ants. The insets provide the results for all the iterations, showing  higher values for both $Bestfval$ and $MRE$. In all cases $\lambda=10$ and $\omega=10.0$. }
    \label{fig:evapQ}
\end{figure}

First we evaluate the influence of the number of ants, the pheromone deposit $Q$ and the evaporation rate on the $Bestfval$ and $MRE$ results for the $5D$ optimization. Figure \ref{fig:evapQ}(a) shows that at $\rho=0$, both $MRE$ and $Bestfval$ are high, indicating that the lack of pheromone evaporation limits the exploration of new paths and compromises the algorithm's performance. For $\rho \geq 0.2$, we observe a general improvement, with stabilized $Bestfval\sim 10$ and small errors $MRE<1.0\%$, when compared to the original FVC error $MRE=2.4\%$. This then indicates that moderate to intense levels of evaporation are essential to preserve the balance between exploration and pheromone deposition. 

Figure~\ref{fig:evapQ}(b) reveals unstable results for $Q<40$, what could then compromise the efficiency of the algorithm, especially for higher number of ants. This highlights the need to reinforce pheromone trails through good solutions, which is directly controlled by $Q$. For $Q>40$, there is a substantial improvement in convergence, towards stability, suggesting relative robustness once this parameter reaches a certain threshold. For 10 ants, although the $MRE$ presents the lowest value for $30 <Q< 70$, it presents significant oscillations, indicating that the algorithm explores the solution space without a clearly defined direction. Thus, although good solutions are eventually found, the algorithm tends to get lost, returning to regions of lower performance. This behavior then reflects the inefficient pheromone deposition, possibly caused by an unstable balance between the amount of pheromone and the number of ants. On the other hand, the configuration with 15 ants, although presenting slightly higher $MRE$ values, demonstrates greater stability throughout the variation of $Q$, with a nearly constant curve for $Q>30$. This stability suggests that the algorithm was able to identify and maintain an optimal solution more consistently. 

Interestingly, the 5 ants optimization returned the lowest $Bestfval$ for most of the $\rho$ and $Q$ regimes, but not the lowest $MRE$. This reflects the fact that, with only 5 ants, the search space is explored in a more restricted and random manner in each iteration, which may cause the algorithm to prematurely converge to a local optimum. Although the numerical differences between 5, 10 and 15 ants for  $Bestfval$ and $MRE$ are small, the results with 15 ants stand out for their consistency for all the parameters. We performed tests with a higher number of ants (not shown), but there was no further improvement. Therefore, 15 ants was adopted here as the upper limit for the number of ants.

We also conducted similar tests (not shown), varying the parameters $\lambda$ (pheromone influence) and $\omega$ (heuristic information), finding stable and very similar results ($MRE\sim 0.8\%$, $Bestfval\sim 10.0$) for $\lambda \geq 2$ and $\omega\geq 2$. For smaller values of $\lambda$ or $\omega$ the performance becomes poorer: for example, for 10 ants with $\lambda=1$ and $\omega=10.0$, we found $MRE\sim7.0\%$ and $Bestfval\sim1500$; while for $\omega=1$ and $\lambda=10.0$, $MRE\sim20.0\%$ and $Bestfval\sim12500$. This degradation highlights the importance of both the pheromone and the heuristic information in guiding the search, particularly for the initial iterations. 

Thus, we explore dimensionality via $Bestfval$ and $MRE$ strategically choosing the optimal ACO parameters: 15 ants, $Q=50$, $\lambda=8.0$, $\omega=8.0$, and $\rho=0.3$. Naturally the optimization time increases with the dimension. The computational cost for a single set of ACO parameters, on a machine using Multi-Threads with 8 Threads at a frequency of $5.80GHz$, were estimated to be: $\sim 110s$ for $1D$; $\sim 125s$ for $2D$; $\sim 140s$ for $3D$; $\sim 147s$ for $4D$ and $\sim 160s$ for $5D$. Since the calculation time increases almost linearly with the dimensionality, the associated cost does not constitute a major limitation for high-dimensional optimizations.

\begin{figure}[htp!]
    \centering
    \includegraphics[width=1\linewidth]{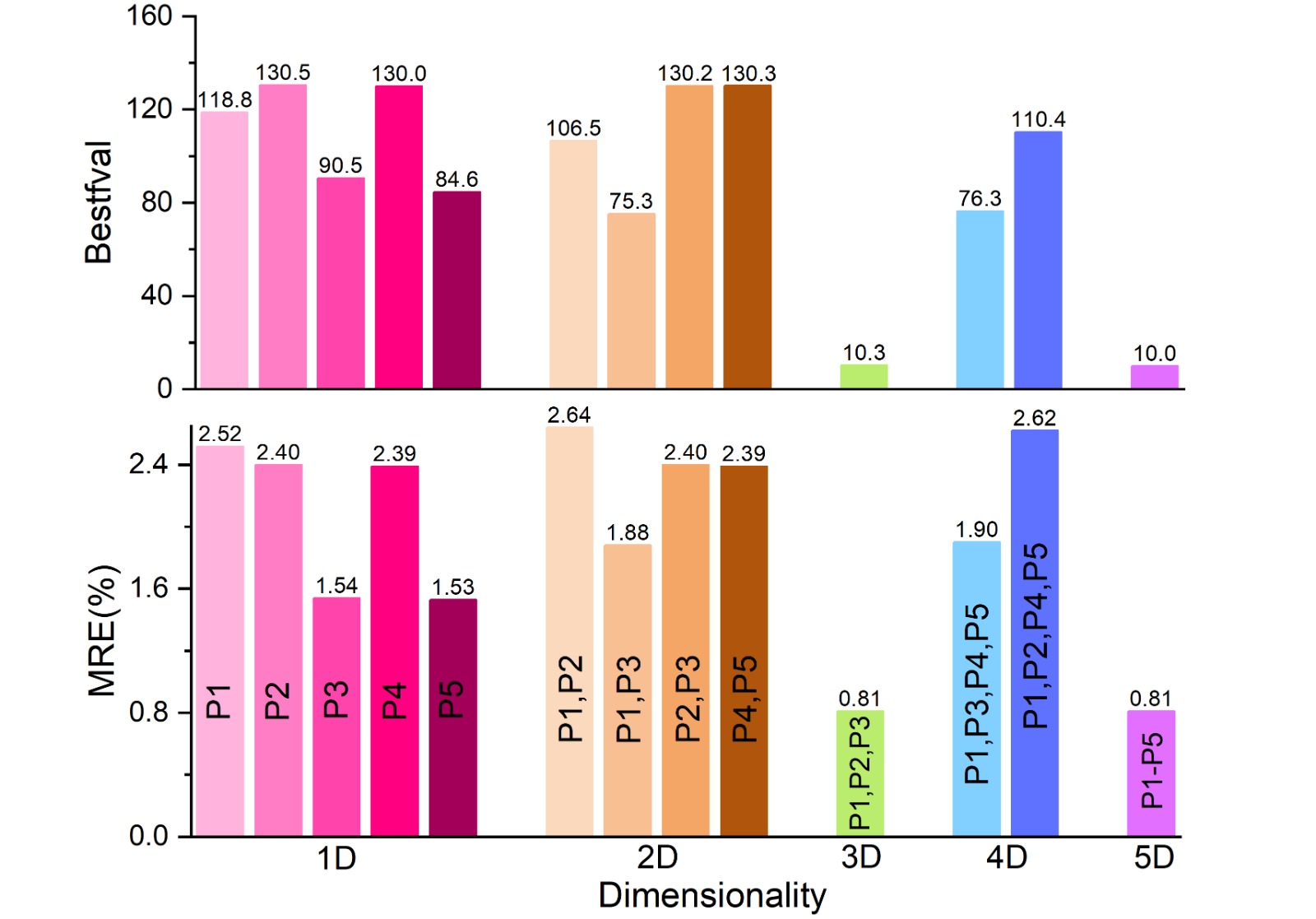}
    \caption{Optimal function value ($Bestfval$) and mean relative error ($MRE$) as a function of the dimensionality of the optimization, for distinct combinations of the optimized parameters $P$ in Eqs. (\ref{eq:alpha}) and (\ref{eq:gama}). All cases were optimized with 1000 iterations, for 15 ants, $Q=50.0$, $\lambda=8.0$, $\omega=8.0$, and $\rho=0.3$.}
    \label{fig:enter-label}
\end{figure}

Figure~\ref{fig:enter-label} shows that for lower dimensions ($1D$, $2D$) and for $4D$, the $Bestfval$ stabilizes at higher values ($Bestfval \geq 75$), suggesting limited exploration of the search space. In contrast, the $3D$ and $5D$ optimizations stabilize at much lower values ($Bestfval\sim 10$). This trend is consistent with the $MRE$ analysis: $3D$ and $5D$ yield the lowest errors ($MRE\sim 0.8\%$), while in other dimensions the $MRE$ remains above $1.5\%$. Nevertheless, in all dimensions, at least one configuration achieves a $MRE$ lower than that of the original FVC functional ($MRE = 2.4\%$). Among them, the $3D$ optimization (providing the optimized parameters $P_1=1.59, P_2=1.44, P_3=0.01$) stands out for offering the best balance between computational cost and performance, achieving a $67\%$ reduction in the original FVC error. Our results then demonstrate that the ACO approach can be successfully used to optimize density functionals, and that the $3D$ optimization offers a better combination of stability, performance and computational time for the optimization of the FVC functional. 

\section{Conclusions}

In this work, we investigated the Ant Colony Optimization (ACO) method and its applicability as an effective alternative to parameter optimization of density functionals, particularly in scenarios where the dimensionality exceeds $2D$.

We conducted a series of tests to evaluate the influence of pheromone update and evaporation parameters on the optimization process, using different numbers of ants ($5$, $10$, and $15$). We could then identify the optimal configurations for these parameters, thus maximizing optimization performance and minimizing the mean relative error ($MRE$) of the ACO functional. Our results revealed that a pheromone evaporation above $0.2$, combined with $15$ ants, yields stabilized $Bestfval$ with the lowest $MRE$. 

With the parameters calibrated, we applied the ACO algorithm to analyze its performance across various dimensionalities. Notably, the $3D$ and $5D$ optimizations exhibited the highest efficiency, achieving a $67\%$ reduction in the error of the original FVC functional. Given that simulation time scales linearly with the optimization dimension, $3D$ optimization offers an effective compromise between accuracy and computational efficiency. Thus, we demonstrate that the ACO method is a robust and promising strategy for complex optimization problems, establishing it as a valuable tool for optimizing density functionals in high-dimensional spaces.
\section*{Acknowledgements}
GMT, RMS, and VVF are supported by São Paulo Research Foundation Fapesp (Grant Nos. $2023$/$17828$-$3$, $2024$/$10789$-$5$, $2021$/$06744$-$8$).
TP and VVF are supported by National Council of
Technological and Scientific Development CNPq (Grant Nos. $140854/2021-5$, $403890/2021-7$, $140854$/$2021$-$5$).

\section*{Author Contributions}
GMT implemented the ant algorithm and performed the functional optimizations. TP performed functional optimizations. All authors analyzed the data and wrote the manuscript.

\end{document}